# Active Control of Ballistic Orbital Transport

*Sobhan Subhra Mishra, James Lourembam, Dennis Jing Xiong Lin, Ranjan Singh\**

S. Mishra, Prof. R. Singh
Division of Physics and Applied Physics, School of Physical and Mathematical Sciences, Nanyang Technological University, 21 Nanyang Link, Singapore 637371, Singapore
E-mail: ranjans@ntu.edu.sg

S. Mishra, Prof. R. Singh
Center for Disruptive Photonic Technologies, The Photonics Institute, Nanyang Technological University, Singapore 639798, Singapore
E-mail: ranjans@ntu.edu.sg

Dr. J. Lourembam, D.J.X. Lin
Institute of Materials Research and Engineering A*STAR (Agency for Science, Technology and Research), 2 Fusionopolis Way, Innovis, #08-03, Singapore 138364, Singapore

**Abstract**

Orbital current, defined as the orbital character of Bloch states in solids, can ballistically travel with larger coherence length through a broader range of materials than its spin counterpart, facilitating a robust, higher density and energy efficient information transmission. Hence, active control of orbital transport plays a pivotal role in propelling the progress of the evolving field of quantum information technology. Unlike spin angular momentum, orbital angular momentum (OAM), couples to phonon angular momentum (PAM) efficiently via orbital-crystal momentum (L-k) coupling, giving us the opportunity to control orbital transport through crystal field potential mediated angular momentum transfer. Here, leveraging the orbital dependant efficient L-k coupling, we have experimentally demonstrated the active control of orbital current velocity using THz emission spectroscopy. Our findings include the identification of a *critical energy density* required to overcome collisions in orbital transport, enabling a swifter flow of orbital current. The capability to actively control the ballistic orbital transport lays the groundwork for the development of ultrafast devices capable of efficiently transmitting information over extended distance.

**Introduction**

The interconversion of spin current and charge current has been extensively explored in the field of spintronics research[1,2]. Notable instances of this conversion encompass phenomena such as spin Hall[2] and inverse spin Hall effects[3,4] observed in heavy metals, spin momentum locking in topological insulators[5,6], and the Rashba-Edelstein effect along with its inverse counterpart in two-dimensional electron gases[7–9]. In recent times, these effects have garnered substantial traction as a potent method for harnessing the spin angular momentum carried by electrons[10,11], allowing for generation of ultrafast charge current, thereby enabling the emission of broadband THz pulses[12,13].

Recent research has brought to light the significance of orbital angular momentum of electrons, giving rise to the emerging field of orbitronics[14,15]. It can possess a substantially greater value compared to its spin counterpart[16]. Additionally, the conversion between orbital and charge current do not require a heavy metal[17,18], thereby expanding the range to include light metals available for utilization. One of the limitations of these systems is the absence of direct source of orbital current; however, this limitation can be eliminated by various orbital pumping techniques through magnetization and lattice dynamics[19]. One of the most used methods of generating orbital current is by using a high Spin Orbit coupling (SoC) ferromagnet like Ni or Co to facilitate the interconversion of spin current and orbital current within the ferromagnet making them indirect sources of orbital current[14–16]. The converted orbital current transports to the adjacent nonmagnetic metal layer where it converts to an accelerated charge current generating THz waves as shown Fig 1(a)[20,21]. Unlike spin transport, which is super diffusive in nature[22], orbital transport is ballistic and hence can propagate over long distance reaching up to 80 nm[21]. Recent study has proposed that the orbital current velocity can be measured up to 0.14 nm/fs in Ni/W heterostructures[21].

Here, we propose an active method to control the orbital current velocity through the applied optical fluence in a Ni/Pt heterostructure. Our findings demonstrate that upon ultrafast photoexcitation, the Ni/Pt heterostructure emits THz radiation primarily due to ballistic transport of orbital current within the Pt layer, which is subsequently converted to charge current. Harnessing electron-phonon coupling which is dependent on orbital angular momentum but independent of spin, we can actively control the orbital transport through laser fluence. Our findings show tunable orbital current velocity from 0.14 nm/fs to 0.18 nm/fs exhibiting a direct control of velocity of orbital transport. We also determine the *critical energy density* necessary to overcome collisions, thereby enabling swifter movement of orbital current within the metal layer.

**Results and Discussion**

While spin and orbital angular momentum have similar symmetry properties, they exhibit distinct behaviors in ultrafast timescales[21,23,24]. Photoexcitation creates spin accumulation $\mu_S$[25,26] which is proportional to the difference in instantaneous magnetization and equilibrium magnetization resulting in release of spin current $j_S$ at a rate proportional to $\mu_S$. In a similar way photoexcitation can also induce orbital accumulation $\mu_L$ resulting in orbital angular momentum transport proportional to $\mu_L$. As the S- type and L- type ultrafast magnetization dynamics is similar for ferromagnets (FM) like Ni[27,28], $\mu_S$ and $\mu_L$ will also be proportional to each other resulting in linear relation between $j_S$ and $j_L$. However, despite identical driving dynamics, spin and orbital current evolution can vary with thickness and time in the nonmagnetic metal (NM) layer as the transport of both the currents can be different at the interfaces and in the bulk of NM. Additionally, due to applied electric field perturbation because of laser fluence, the angular momentum exchange between the lattice and orbital wave function can also control the transport of orbital angular momentum.[19]

As depicted in Figure 1(a), when the FM/NM heterostructure is subjected to ultrafast photoexcitation, the initially generated ultrafast spin current partially transforms into an ultrafast orbital current due to LS correlation in ferromagnet layer resulting in the injection of both spin $(j_S)$ and orbital current $(j_L)$ into the nonmagnetic metal layer. Through the processes of LCC and SCC, accelerated charge current is induced, serving as the source of THz radiation. The resultant THz electric field is directly proportional to the sheet charge current $(I_C(t))$, as described by the following equation[12,13,29].

$$E(t) \propto I_C(t) \propto \int_{-d_{FM}}^{d_{NM}} dz [\theta_{LC}(z) j_L(z,t) + \theta_{SC}(z) j_S(z,t)] \quad (1)$$

Here $d_{FM}$ and $d_{NM}$ denotes the thickness of the ferromagnet and nonmagnetic metal layer respectively and $\theta_{LC}$ and $\theta_{SC}$ denote the orbital Hall and spin Hall angle which govern the efficiency of orbital to charge (LCC) and spin to charge conversion (SCC). The equation does not include the minor process of THz emission like ultrafast demagnetization and anomalous hall effect due to the ferromagnet as those can be separated out from the photocurrent mechanisms experimentally[30].

Relative sign of spin hall and orbital hall angle depends on the LS correlation $\langle L.S \rangle$ as shown in Fig 1(b) and 1(c). If L.S <0, the spin and orbital transport will have opposite sign whereas if L.S>0, spin and orbital transport will have same sign[31]. To facilitate a direct comparison between orbital transport and spin transport, we study two different types of FM/NM heterostructures. In the first case, Ni was chosen as the FM layer to illustrate orbital transport as Ni has higher efficiency of generating orbital current from spin current because of its higher L.S correlation near fermi level[32]. In the second case, to demonstrate spin transport, NiFe is chosen as FM due to its higher spin current generation efficiency[12]. In both the cases, NM chosen was Pt due to its high Orbital Hall Conductivity and Spin Hall Conductivity[33]. An in-plane constant magnetic field of 128 mT was applied to keep the system at saturated magnetization state.

Figure 2(a) displays the emitted THz radiation from a Ni(3 nm)/Pt(x nm) heterostructure, when photoexcited by a constant fluence of 1270 µJ/cm$^2$, with variable Pt thickness (x = 3, 6, 9, 18). It is observed that as the Pt thickness increases, the emitted THz pulse encounters a delayed arrival, whereas in case of NiFe (3 nm)/ Pt (x nm) with x = 1, 2, 4, 6, 8, there is no delay in THz pulse as illustrated in Figure 2(b). This comparison serves to establish the prevalence of different photocurrent mechanism of THz emission from Ni/Pt compared to NiFe/Pt. In case of NiFe/Pt where spin transport dominates the THz emission, the increase in Pt thickness does not delay the arrival of THz due to lower relaxation length of spin current. A very small delay of around 2 fs can be induced (See supplementary for calculations) due to the THz refractive index of platinum. Therefore, the delay in Ni/Pt could be attributed to a long-distance transport of orbital angular momentum.

The shift in the peak THz pulse is graphically represented in Figure 2(c), demonstrating the linear correlation between Pt thickness and the delay obtained in THz peak which validates the ballistic nature of orbital transport and provides us with an orbital current velocity of approximately 0.18 nm/fs, as opposed to the zero delay in THz peak value when increasing the thickness of Pt in the case of NiFe/Pt, where spin transport predominates. Furthermore, due to its larger relaxation length, orbital current will experience more significant angular dispersion at higher thicknesses, resulting in the widening of the THz pulse depicted in Figure 2(d). The pulse width measured from peak-to-peak time increases at larger thickness of Pt in case of Ni/Pt. However, in case of NiFe/Pt where spin transport dominates, due to the very short relaxation time, the spin current does not travel enough distance to have large angular dispersion resulting in constant pulse width with increase in thickness of Pt.

The orbital angular momentum of the nonlocalized electrons interacts with lattice through the crystal field potential with following continuity equation[34,35]

$$\frac{\partial}{\partial t}\langle L \rangle = \langle F^L \rangle + \frac{1}{i\hbar}\langle [L, V_{CF}] \rangle + \langle \lambda S \times L \rangle \quad (2)$$

Here $F^L$ is the orbital flux term, $\langle [L, V_{CF}] \rangle$ describes the transfer of angular momentum between orbital and crystal, $V_{CF}$ is crystal field potential and $\lambda S \times L$ describes the mutual transfer of spin and orbital angular momentum within a single electron. The electric field of the applied fluence interacts with the orbital, causing a perturbation in the orbital wave function which gives rise to a non-equilibrium orbital wave function. The electric field perturbed non-equilibrium orbital wave function extracts angular momentum from the lattice, thereby increasing the velocity of orbital transport.

Figure 3 offers a comprehensive illustration of the active control of orbital transport within the Pt layer. The thickness of the ferromagnetic material Ni remains constant at 3 nm throughout the experiment. In Figures 3(a-d), the emitted THz pulse is depicted for Ni(3 nm) and Pt(x nm), where x takes values of 3, 6, 9, and 18, respectively, at different fluence levels. As the fluence increases, a noticeable shift in the emitted THz pulse is observed. Initially, the pulse shifts towards the right, indicating a delayed arrival of the THz pulse. However, beyond a certain fluence threshold referred to as the *critical fluence*, there is an increase in orbital current velocity. Consequently, beyond this *critical fluence*, there is left shift in the arrival time of the THz pulse. Similar behaviour was also observed in other orbital transport-based emission systems such as Ni/Ru (See supplementary section). Figure 3(e) demonstrates that critical fluence is contingent on the thickness of the nonmagnetic Pt layer. As the thickness is increased, the carriers require more energy to overcome collisions to move ballistically over a larger distance, resulting in a higher critical fluence. Similar behaviour was also observed when the Peak-to- peak time was recorded as shown in Fig 3(f).

Figure 4(a) elucidates the schematic representation of the orbital transport mechanism both prior to and post reaching critical fluence. Non localized electrons, bearing information about orbital angular momentum, traverse through orbital hopping between nuclei within a solid. An increase in fluence increases the number of charge carriers thereby enhancing the number of collisions that impede the transport process. Nevertheless, beyond the critical fluence, nonlocalized electrons perturbed by laser fluence absorb additional angular momentum from the lattice according to equation 2[19,34,35], as illustrated in Figure 4(b). As we apply laser fluence, the magnetic moment of localized electrons couples with the nonlocalized electron through exchange interaction[10], thus creating a spin current. Due to high spin orbit correlation of Ni near Fermi level, there is angular momentum transfer between spin and orbital which can be explained by the cross product of S and L ($\langle \lambda S \times L \rangle$) as shown in equation 2. Additionally due to the electric field of the applied laser fluence, a perturbation in the orbital wave function is induced, thus creating a non-equilibrium state. Consequently, the non-equilibrium orbital wave function takes angular momentum from the lattice through the crystal field potential $V_{CF}$ explained by $\frac{1}{i\hbar}\langle[L, V_{CF}]\rangle$ in the equation 2, enabling the orbital current to surpass collisions, facilitating a more rapid transport.

Evidently, the critical fluence is linearly corelated with the thickness of the heavy metal layer. Figure 4(c) illustrates that as the thickness increases, the critical fluence also increases proportionally. The slope of the linear relation can be called as *critical energy density* denoted by $\varepsilon_C$ and quantified as 343 J/cm$^3$ in case of Ni/Pt heterostructure representing the energy required per unit volume to overcome the collision and facilitate a complete ballistic orbital transport in Pt. For different orbital converters, $\varepsilon_C$ can be different depending on the intrinsic properties of the materials, and further investigation on this is required. Finally, the orbital current velocities at different fluences were extracted by recording the delay in the THz peak with respect to the Ni (3 nm)/ Pt (3 nm) heterostructure. Given the ballistic nature of orbital transport, the slope of the linear relationship between delay and the increase in heavy metal thickness was utilized to calculate the velocity. As shown in Fig 4(d), the orbital current velocity was extracted at 191 μJ/cm$^2$ and found out to be 0.14 nm/fs. As we increase the fluence, the slope started increasing and at 382 μJ/cm$^2$ the orbital current velocity was 0.16 nm/fs and at 1270 μJ/cm$^{2,}$ the velocity was found to be 0.18 nm/fs. The linear relation also proves the ballistic nature of the orbital transport in Pt layer over a larger distance than spin transport.

In summary, leveraging THz emission spectroscopy, we demonstrate THz emission from optically excited Ni/Pt heterostructure predominantly from long range ballistic orbital transport. Furthermore, the orbital transport can be controlled through the electric field of the applied fluence. Absorption of higher energy initially leads to more charge carrier formation enhancing the collision thus delaying the transport. However, after a critical fluence, carriers overcome the collision and the orbital transport happen at a swifter pace due to absorption of phonon angular momentum. Exploiting this phenomenon, we have also highlighted the active enhancement of experimentally calculated orbital current velocity from 0.14 nm/fs to 0.18 nm/fs through increase in fluence enabling longer and faster orbital transport. Our findings establish an approach to control the long-distance ballistic L transport, thus creating new opportunities to design future ultrafast devices with orbitronics materials. Additionally, because of the orbital dependent efficient phonon orbital coupling, it is also possible to have lattice assisted orbital pumping called Orbital Angular Position (OAP)[19,36]. Thus, integrating twistronics and orbitronics towards THz emission, we envision an Orbitronic Terahertz Emitter (OTE) without the application of magnetic field.

## 4. Methods

*Sample preparation*: The FM/NM films were deposited on 1 mm- Quartz substrates by d.c. magnetron sputtering at room temperature, using a Chiron ultrahigh vacuum system with a vacuum base pressure of 1× 10−8 torr.

*THz emission experiment*: The THz radiation emitted is captured using a 1 mm thick ZnTe crystal oriented along the <110> axis, known for its nonlinear properties. The femtosecond laser pulse, which illuminates the orbitronics heterostructure, has a wavelength of 800 nm, corresponding to a laser energy of 1.55 electron volts (eV). It has a pulse width of 35 femtoseconds (fs) and operates at a repetition rate of 1 kHz. A beam splitter is employed to divide it into two parts. The higher intensity portion is directed towards photoexcitation of the emitter, while the lower intensity portion serves as a probe for detection. Precise time matching is ensured by a mechanical delay stage. The details about the THz emission spectroscopy set up can be found in supplementary section. As the emitted THz pulse, collected by parabolic mirrors, focuses on the ZnTe <110> detector, it induces birefringence in the crystal. Simultaneously, the time-matched probe beam traverses through the crystal and encounters a change in its polarization, directly proportional to the birefringence. For the electro-optic detection[37], a quarter-wave plate and a Wollaston prism are employed to distinguish the s and p polarization of the laser. The rotational changes in the probe laser are subsequently identified using a balanced photodiode that measures the intensity difference between the s and p polarized light. The electrical signal thus detected undergoes initial pre-amplification before being input into the lock-in amplifier to enhance the signal-to-noise ratio. The resultant signal from the lock-in amplifier is utilized to generate the THz electric field through a homemade LabVIEW code.

**Data availability**

The data supporting this study's findings are available from the corresponding author upon reasonable request. Supplementary information is linked to the online version of the paper.

**Acknowledgments**

The authors thank Thomas Tan and Baolong Zhang for valuable discussions and suggestions. R.S. and S.M. would like to acknowledge the Ministry of Education (MoE), Singapore, for the support through MOE-T2EP50121-0009. J.L. and D.J.X.L would like to acknowledge funding support from SpOT-LITE programme (Agency for Science, Technology and Research, A*STAR Grant No. A18A6b0057) through RIE2020 funds from Singapore.

**Author Contributions**

SM, J.L and R.S. conceived the project. S.M and R.S designed the experiments. S.M. performed all the THz measurements and experimental analysis. D.J.X.L and J.L. fabricated the orbitronic emitter. All the authors analysed and discussed the results. S.M. and R.S. wrote the manuscript with inputs from J.L. R.S. lead the overall project.

**Author Information**

Reprints and permissions information is available online. The authors declare no competing financial interests. Correspondence and requests for materials should be addressed to Prof. Ranjan Singh (ranjans@ntu.edu.sg).

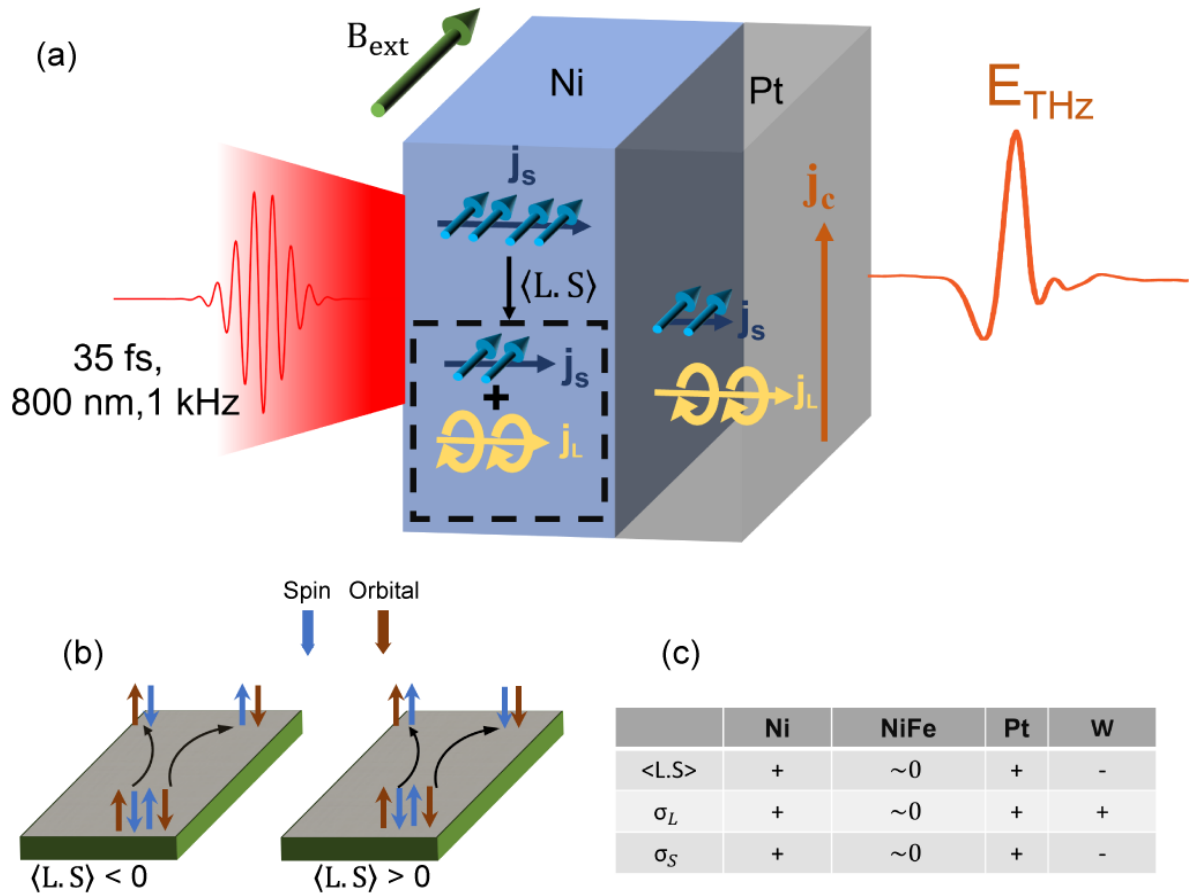

**Figure 1: Pumping and Detection of Terahertz Orbital (L) and Spin(S) currents** (a) Upon ultrafast photoexcitation of the FM, spin currents are generated and partially converted to orbital currents due to L.S correlation, injecting both spin current ($j_S$) having shorter relaxation length and orbital current ($j_L$) having longer relaxation length into metal layer. Orbital-to-charge conversion (LCC) and spin-to-charge conversion (SCC) processes generate an accelerated charge current ($j_C$) to emit a THz radiation. (b) L.S correlation dependance of direction of orbital current and spin current propagation in Orbital Hall effect (c) Sign of spin orbit coupling and the orbital and spin hall conductivity for Ni, NiFe, Pt and W.

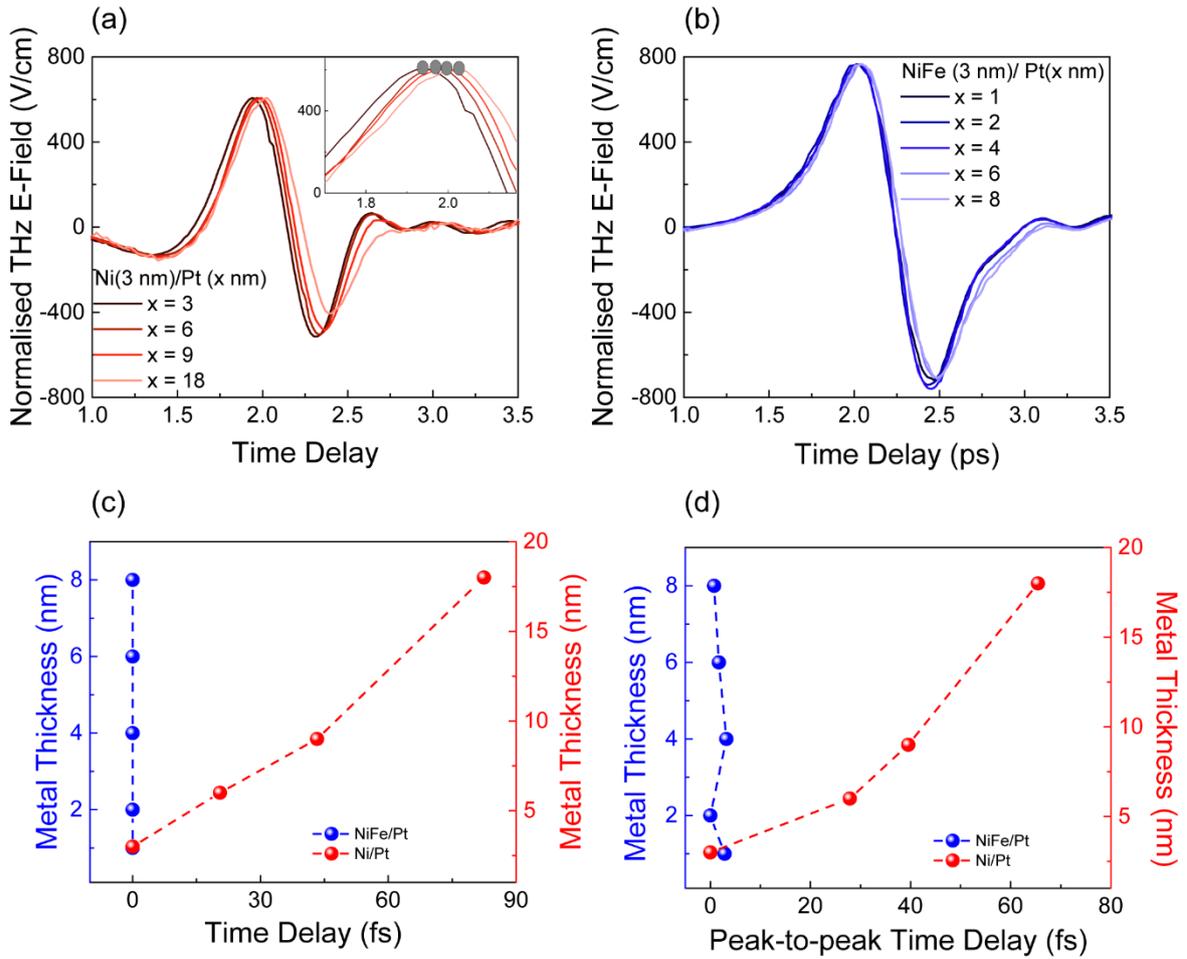

**Figure 2: Ballistic Orbital Transport in Ni/Pt heterostructure** (a) Terahertz signal emitted from Ni(3 nm) /Pt(x nm) heterostructure with variable Pt thickness x = 3, 6 , 9 and 18; Inset shows the zoomed in peak of the THz pulse indicating a right shift as we increase the Pt thickness (b) Terahertz signal emitted from NiFe (3 nm) /Pt (x nm) heterostructure with variable Pt thickness with x = 1, 2, 4, 6 and 8 indicating no right shift in the peak. (c) Delay in the emitted THz pulse with increase in Pt thickness demonstrating the ballistic orbital transport in Ni/Pt heterostructure in contrast to spin transport in NiFe/Pt heterostructure (d) Increase in peak-to-peak width indicating widening of the emitted THz pulse with increase in Pt thickness in Ni/Pt heterostructure (Orbital Transport) providing evidence of the increase in the angular dispersion proving the longer relaxation length of orbital current in contrast to spin transport in NiFe/Pt heterostructure

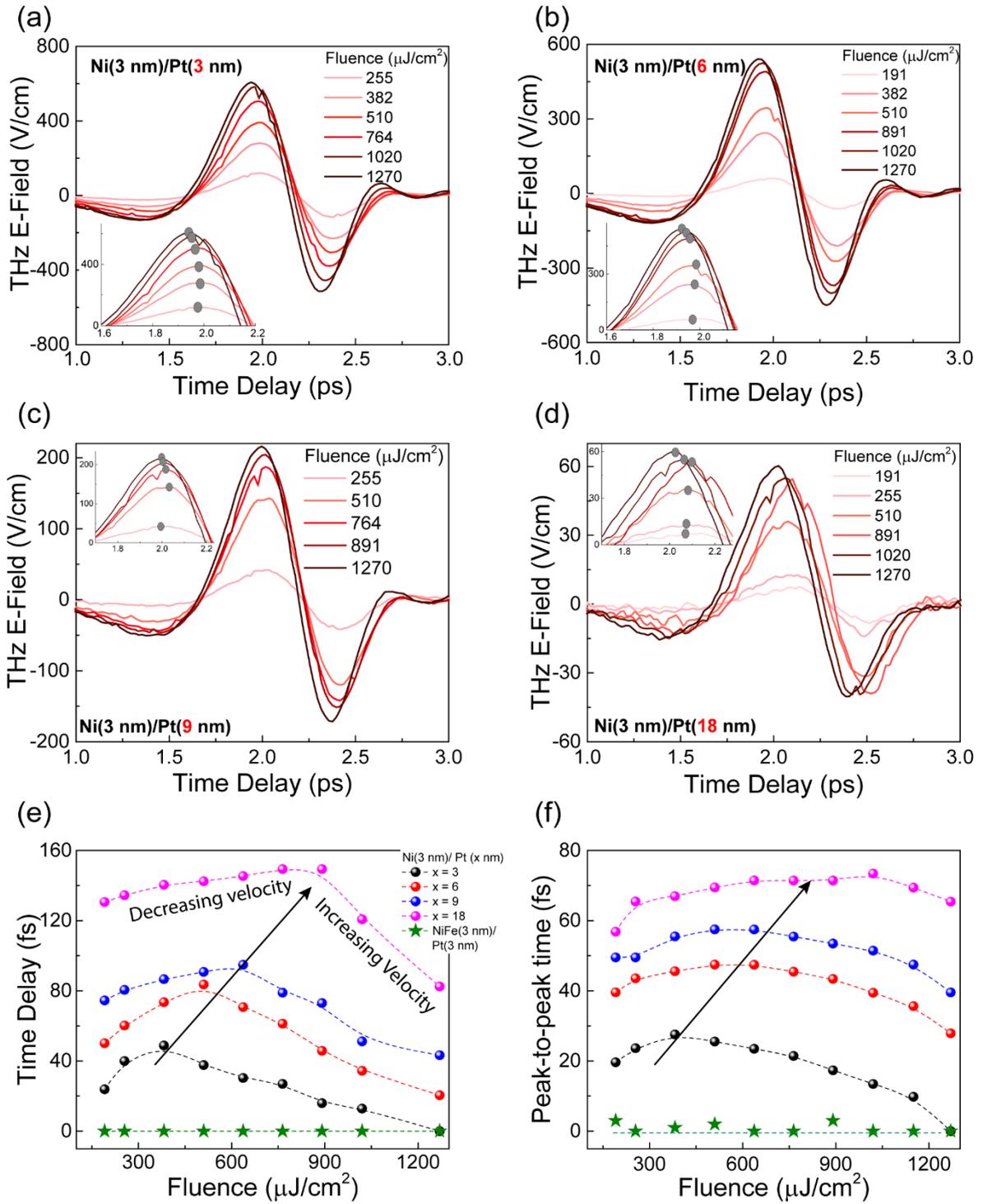

**Figure 3: Active Control of Ballistic Orbital Transport.** THz emission from Ni (3nm)/Pt (x nm) at different fluence when (a) x = 3 nm, (b) x = 6 nm, (c) x = 9 nm, (d) x = 18 nm, A clear shift in THz peak was observed as we change the fluence Extracted (e) Time delay for different fluence is shown for Ni (3 nm)/ Pt (x nm) with x = 3, 6, 9, 18. Initially the shift is towards right

indicating the decrease in orbital velocity and after a critical fluence, delay starts decreasing with increase in fluence showing swifter orbital transport; Similar shift is not seen in spin transport as shown in NiFe (3 nm)/ Pt (3 nm). (f) Peak-to-peak time difference of THz pulse, for different fluence is shown for Ni (3 nm)/ Pt (x nm) with x = 3, 6, 9 and 18. Similar behavior as (e) can be seen

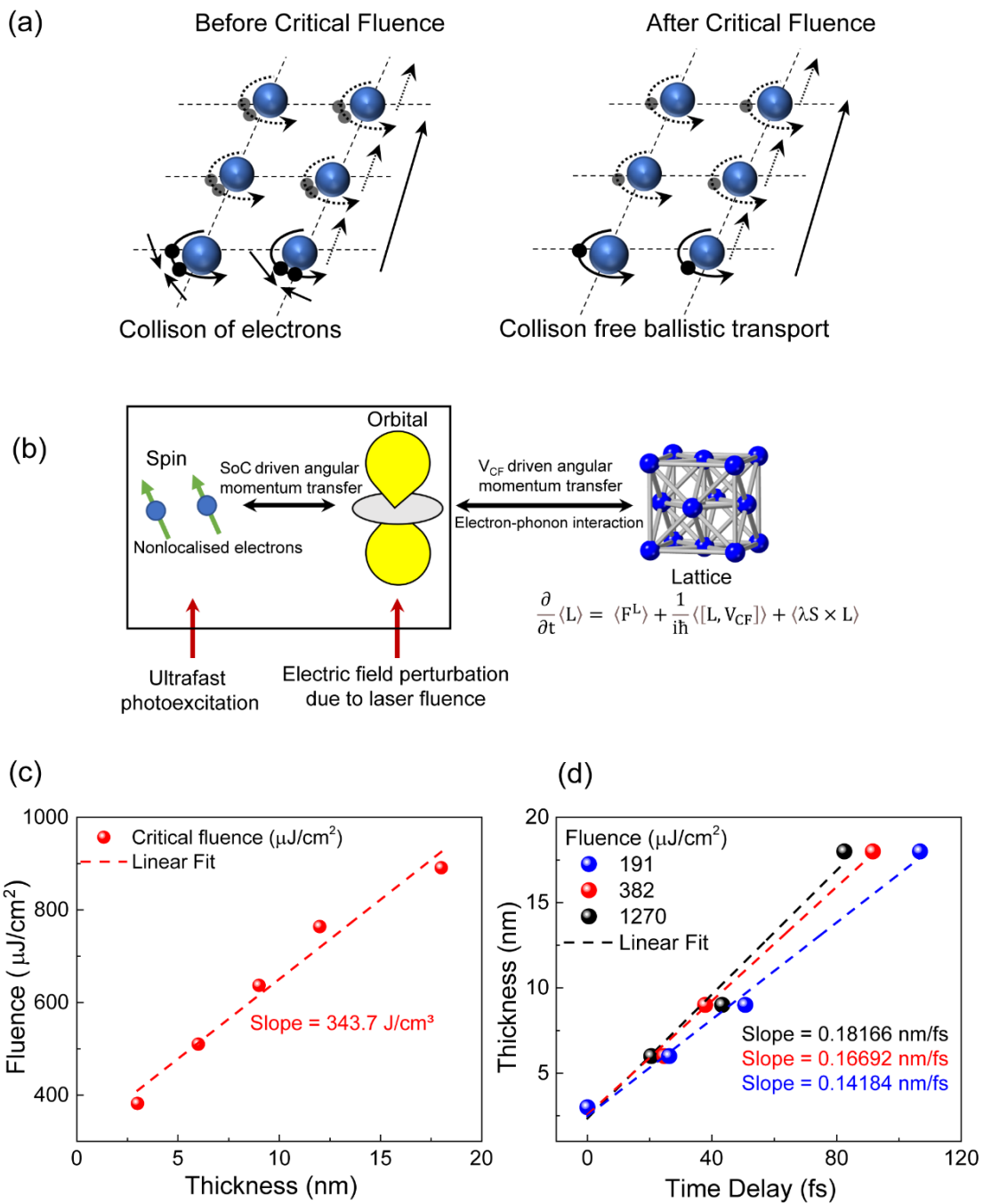

**Figure 4: Active Control of Orbital Current Velocity**. (a) Schematic of transport of orbital angular momentum, before critical fluence collision between charge carriers make the transport slower, after critical fluence, swifter transport of orbital angular momentum takes place (b) Mechanism of increase in orbital current velocity due to angular momentum transfer between lattice and electrons. c) Extracted critical fluence with variation in thickness. The slope of the fitted straight line indicates the "critical energy density" (d) Extracted time delay with thickness at different fluences. The slope of the linear fitted line can be termed as orbital current velocity which could be tuned from 0.14 nm/fs to 0.18 nm/fs as the fluence was increased from 191 µJ/cm$^2$ from 1270 µJ/cm$^2$

**References**

1. Wolf, S. A. *et al.* Spintronics: A Spin-Based Electronics Vision for the Future. *Science* **294**, 1488–1495 (2001).

2. Hirsch, J. E. Spin Hall effect. *Phys. Rev. Lett.* **83**, 1834–1837 (1999).

3. Miao, B. F., Huang, S. Y., Qu, D. & Chien, C. L. Inverse Spin Hall Effect in a Ferromagnetic Metal. *Phys. Rev. Lett.* **111**, 066602 (2013).

4. Saitoh, E., Ueda, M., Miyajima, H. & Tatara, G. Conversion of spin current into charge current at room temperature: Inverse spin-Hall effect. *Appl. Phys. Lett.* **88**, 182509 (2006).

5. Luo, S., He, L. & Li, M. Spin-momentum locked interaction between guided photons and surface electrons in topological insulators. *Nat. Commun.* **8**, 2141 (2017).

6. Zhang, H., Liu, C.-X. & Zhang, S.-C. Spin-orbital Texture in Topological Insulators. *Phys. Rev. Lett.* **111**, 066801 (2013).

7. Schliemann, J. & Loss, D. Anisotropic transport in a two-dimensional electron gas in the presence of spin-orbit coupling. *Phys. Rev. B* **68**, 165311 (2003).

8. Zhou, C. *et al.* Broadband Terahertz Generation via the Interface Inverse Rashba-Edelstein Effect. *Phys. Rev. Lett.* **121**, 086801 (2018).

9. Comstock, A. *et al.* Spintronic Terahertz Emission in Ultrawide Bandgap Semiconductor/Ferromagnet Heterostructures. *Adv. Opt. Mater.* **11**, 2201535 (2023).

10. Agarwal, P. *et al.* Terahertz spintronic magnetometer (TSM). *Appl. Phys. Lett.* **120**, 161104 (2022).

11. Agarwal, P., Huang, L., Ter Lim, S. & Singh, R. Electric-field control of nonlinear THz spintronic emitters. *Nat. Commun.* **13**, 4072 (2022).

12. Seifert, T. *et al.* Efficient metallic spintronic emitters of ultrabroadband terahertz radiation. *Nat. Photonics* **10**, 483–488 (2016).

13. Kampfrath, T. *et al.* Terahertz spin current pulses controlled by magnetic heterostructures. *Nat. Nanotechnol.* **8**, 256–260 (2013).

14. Lee, S. *et al.* Efficient conversion of orbital Hall current to spin current for spin-orbit torque switching. *Commun. Phys.* **4**, 234 (2021).

15. Li, T. *et al.* Giant Orbital-to-Spin Conversion for Efficient Current-Induced Magnetization Switching of Ferrimagnetic Insulator. *Nano Lett.* acs.nanolett.3c02104 (2023) doi:10.1021/acs.nanolett.3c02104.

16. Go, D., Jo, D., Lee, H.-W., Kläui, M. & Mokrousov, Y. Orbitronics: Orbital currents in solids. *EPL Europhys. Lett.* **135**, 37001 (2021).

17. Sala, G. & Gambardella, P. Giant orbital Hall effect and orbital-to-spin conversion in 3 d , 5 d , and 4 f metallic heterostructures. *Phys. Rev. Res.* **4**, 033037 (2022).

18. Jo, D., Go, D. & Lee, H.-W. Gigantic intrinsic orbital Hall effects in weakly spin-orbit coupled metals. *Phys. Rev. B* **98**, 214405 (2018).

19. Han, S. *et al.* Theory of Orbital Pumping.

20. Wang, P. Inverse orbital Hall effect and orbitronic terahertz emission observed in the materials with weak spin-orbit coupling. *Npj Quantum Mater.* (2023).

21. Seifert, T. S. *et al.* Time-domain observation of ballistic orbital-angular-momentum currents with giant relaxation length in tungsten. *Nat. Nanotechnol.* (2023) doi:10.1038/s41565-023-01470-8.

22. Battiato, M., Carva, K. & Oppeneer, P. M. Superdiffusive Spin Transport as a Mechanism of Ultrafast Demagnetization. *Phys. Rev. Lett.* **105**, 027203 (2010).

23. Bose, A. *et al.* Detection of long-range orbital-Hall torques. *Phys. Rev. B* **107**, 134423 (2023).

24. Hayashi, H., Go, D., Mokrousov, Y. & Ando, K. Observation of orbital pumping.

25. Beaurepaire, E. Ultrafast Spin Dynamics in Ferromagnetic Nickel. *Phys. Rev. Lett.* **76**, (1996).

26. Choi, G.-M., Min, B.-C., Lee, K.-J. & Cahill, D. G. Spin current generated by thermally driven ultrafast demagnetization. *Nat. Commun.* **5**, 4334 (2014).

27. Go, D. *et al.* Orbital Pumping by Magnetization Dynamics in Ferromagnets. Preprint at http://arxiv.org/abs/2309.14817 (2023).

28. Stamm, C., Pontius, N., Kachel, T., Wietstruk, M. & Dürr, H. A. Femtosecond x-ray absorption spectroscopy of spin and orbital angular momentum in photoexcited Ni films during ultrafast demagnetization. *Phys. Rev. B* **81**, 104425 (2010).

29. Liu, Y. *et al.* Separation of emission mechanisms in spintronic terahertz emitters. *Phys. Rev. B* **104**, 064419 (2021).

30. Zhang, W. *et al.* Ultrafast terahertz magnetometry. *Nat. Commun.* **11**, 4247 (2020).

31. Go, D., Jo, D., Kim, C. & Lee, H.-W. Intrinsic Spin and Orbital Hall Effects from Orbital Texture. *Phys. Rev. Lett.* **121**, 086602 (2018).

32. Go, D. Theory of Orbital Hall Effect and Current-Induced Torques by Orbital Current. (2018).

33. Kontani, H., Tanaka, T., Hirashima, D. S., Yamada, K. & Inoue, J. Giant Orbital Hall Effect in Transition Metals: Origin of Large Spin and Anomalous Hall Effects. *Phys. Rev. Lett.* **102**, 016601 (2009).

34. Haney, P. M. & Stiles, M. D. Current-Induced Torques in the Presence of Spin-Orbit Coupling. *Phys. Rev. Lett.* **105**, 126602 (2010).

35. Go, D. *et al.* Theory of current-induced angular momentum transfer dynamics in spin-orbit coupled systems. *Phys. Rev. Res.* **2**, 033401 (2020).

36. Han, S., Lee, H.-W. & Kim, K.-W. Orbital Dynamics in Centrosymmetric Systems. *Phys. Rev. Lett.* **128**, 176601 (2022).

37. Bakker, H. J., Cho, G. C., Kurz, H., Wu, Q. & Zhang, X.-C. Distortion of terahertz pulses in electro-optic sampling. *J. Opt. Soc. Am. B* **15**, 1795 (1998).